\documentclass{pasj00}

\begin{document}
\SetRunningHead{M. Morii et al.}{AXP 1E 1841-045}
\Received{2002/12/17}
\Accepted{2003/03/24}

\title{{\it Chandra} Observation of the Anomalous X-ray Pulsar 1E 1841-045}

\author{Mikio \textsc{Morii}\altaffilmark{1},
	Rie \textsc{Sato}\altaffilmark{1},
	Jun \textsc{Kataoka}\altaffilmark{1}, and
	Nobuyuki \textsc{Kawai}\altaffilmark{1,2}}
\altaffiltext{1}{Tokyo Institute of Technology,
	 2-12-1, Ohokayama, Megro-ku, Tokyo 152-8551}
\email{morii@hp.phys.titech.ac.jp}
\altaffiltext{2}{The Institute of Physical and Chemical Research,
	 2-1, Hirosawa, Wako-shi, Saitama 351-0198}


%


\maketitle

\begin{abstract}
We present the results from the {\it Chandra} ACIS CC mode observation of
an anomalous X-ray pulsar (AXP) 1E 1841-045.
This is the first observation in which
the pulsar spectrum in wide energy range
is spatially discriminated from
the surrounding SNR, Kes 73.
Like other AXPs, the phase-integrated spectrum is fitted well
with power-law plus blackbody model.
The spectral parameters are
$\Gamma = 2.0 \pm 0.3$, $kT_{\rm BB} = 0.44 \pm 0.02$ keV, and
$N_H = 2.54^{+0.15}_{-0.13} \times 10^{22} {\rm cm}^{-2}$.
This photon index is significantly flatter than the other AXPs,
and resemble to soft gamma-ray repeaters (SGRs) in the quiescent state.
The pulse profile is double-peaked,
and we found that the second peak has significantly hard spectrum.
The spectra of all phases are consistent with 
power-law plus blackbody model
with constant temperature and photon index.
When fitted with two blackbody model,
we obtained similarly good fit.
These results can be interpreted that
there are two emission regions
with different energy spectra.
\end{abstract}

\section{Introduction}
The anomalous X-ray pulsars (AXPs) make a small,
but unique group (5 or 6) in the X-ray pulsars ($\sim 100$).
Their characteristics are as follows
(see \cite{Mereghetti 2002} for a review);
(1) much softer X-ray spectra than the
more common high mass X-ray binary pulsars,
(2) the steady spin-down on the timescales of $10^3-10^5$ yrs,
(3) spin periods narrowly range in $\sim5-12$ s.
AXPs are mysterious, since their energy source is still unknown.
First of all, they cannot be rotation-powered, because
observed X-ray luminosities
(${\rm L}_{\rm X} \sim 10^{34}- 10^{36}$ ergs/s) exceed
the spin-down energy loss rates of the neutron stars
 ($\dot{{\rm E}} = 4\pi^2 {\rm I} \dot{{\rm P}}/{\rm P}^3 \sim 10^{32.5}$
 ergs/s).
Secondly, they are unlikely accretion-powered,
because any binary companion stars, which provide the accreting mass to the
neutron stars, have not been detected.

The most plausible and exotic hypothesis is that
AXPs are isolated neutron stars with ultra-strong magnetic field
($\sim 10^{14}-10^{15}$ G), so-called ``Magnetar''
(\cite{Thompson Duncan 1993}).
In this model the X-ray photons are produced by the release of strong
magnetic energy stored in the neutron star crust
(\cite{Thompson Duncan 1996}).
Due to the strong magnetic field
($B \gtrsim B_{\rm cr} = m_e^2 c^3/e \hbar = 4.4 \times 10^{13}$G),
exotic QED processes are predicted:
photon splitting
and vacuum polarization,
etc.

There are another type of magnetar candidate:
soft gamma-ray repeaters (SGRs).
AXPs and SGRs exist on a similar region
in ${\rm P}$$-$$\dot{\rm P}$ diagram.
The persistent X-ray luminosities of SGRs are also much
larger than their spin-down luminosities.
In contrast to AXPs, SGRs show occasional short bursts,
and have relatively harder spectra ($\Gamma \sim 2$)
and stronger magnetic fields than those of AXPs
(\cite{Marsden White 2001}).

1E 1841-045 was discovered as a compact X-ray source centered
in a supernova remnant (SNR) Kes 73 (G27.4+0.0) by {\it Einstein} HRI
\citep{Kriss 1985}.
The kinematic distance toward the SNR was determined
to be between 6 and 7.5 kpc, by using the H I absorption data
obtained with the VLA
(\cite{Sanbonmatsu 1992}).
The X-ray pulsation of $11.8$ s period was discovered with {\it ASCA}
\citep{Vasisht Gotthelf 1997}
and the central source has been identified as an anomalous X-ray pulsar.
Although 1E 1841-045 was observed with {\it Ginga, ROSAT, ASCA, RXTE}
and {\it BeppoSAX}
(\cite{Helfand 1994}, \cite{Gotthelf Vasisht 1997},
\cite{Gotthelf 1999}; 2002),
pulsar component could not be studied separately from 
the surrounding SNR component.
In this paper we report the spatially resolved
spectroscopy of 1E 1841-045 for the first time with {\it Chandra}
thanks to its high spatial resolution
of $\sim 0.5^{\prime \prime}$ (FWHM).

\section{Observation and data reduction}
\label{sec: obs and red}
\begin{figure}
 \begin{center}
    \FigureFile(80mm,80mm){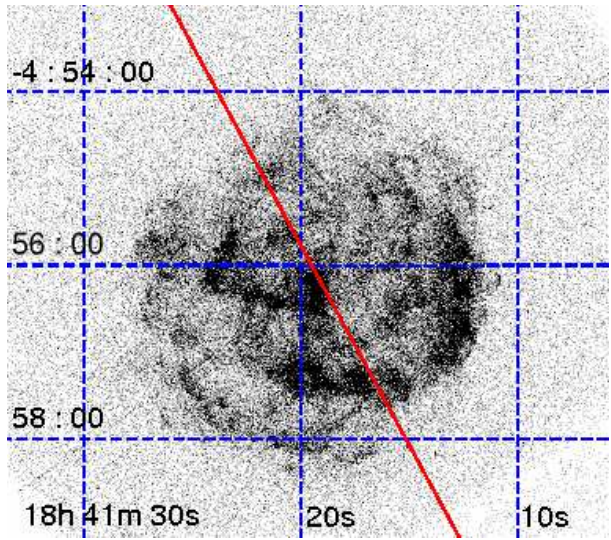}
 \end{center}
  \caption{The image of Kes 73 on the ACIS-S3 chip in TE mode.
	Coordinates are J2000.
	 The projection axis in CC mode is shown in the slanted line.
}
	\label{fig: 2d}
\end{figure}

Kes 73 and 1E 1841-045 at its center were
observed with {\it Chandra}
using the Advanced CCD Imaging Spectrometer (ACIS)
in timed exposure (TE) mode (30ks) on 2000 July 23
and continuous clocking (CC) mode (10ks) on 2000 July 29.
The image of Kes 73 taken in the TE mode
is shown in Figure \ref{fig: 2d}.
The source was placed on the nominal target position of
ACIS-S3, a back-illuminated CCD on the spectroscopic array (ACIS-S)
with good charge transfer efficiency and spectral resolution
(\cite{Townsley 2002}).
In addition to S3, four front-illuminated CCDs were active
(I0, I1, I2, I3, and S2).
The focal plane operating temperature was $-120^{\circ}{\rm C}$.

For the pulsar analysis, we have used the CC mode data
of level 2 product made by the {\it Chandra} X-ray Center.
The net exposure was 10.497 ks.
In CC mode, the pileup effect is negligibly small
because readout is very short (2.85 ms).
It enables us both accurate time-integrated
and pulse-phase resolved spectroscopy of the pulsar,
but the image is degenerated into one-dimensional.

\begin{figure}
 \begin{center}
    \FigureFile(80mm,80mm){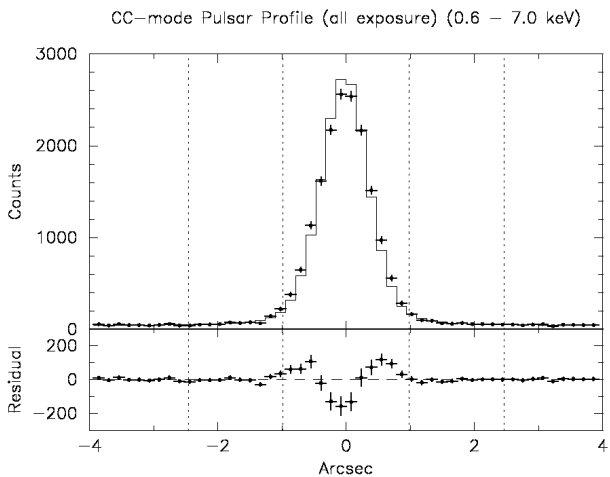}

 \end{center}
  \caption{One-dimensional image of 1E 1841-045
	obtained in CC mode (the upper panel).
	Horizontal and vertical axes are shown in units of
	arcsec and total counts during all exposure, respectively.
	Data and their 1$\sigma$ errors are shown in crosses.
	The boundaries of the source and background regions
	are shown by vertical dotted lines (see text).
	The best fit model consisting of a point source and
	a constant background is shown by a solid histogram.
	The residual from the best-fit model is shown in
	the bottom panel.}
	\label{fig: 1d}
\end{figure}

The projected image around the pulsar
in the energy range of 0.6$-$7.0 keV
is shown in Figure \ref{fig: 1d}.
We selected the source photons from
the 4 pixels width ($\sim 2^{\prime \prime}$)
centered on the peak pixel.
The background has been taken from two segments
adjacent to the source region.
They are net 6 pixels ($\sim 3^{\prime \prime}$)
wide in total, thus 1.5 times larger than the source region.
The background flux amounts to $5.9\%$ of the source flux
in the 0.6$-$7.0 keV energy range.

\section{Timing Analysis} \label{sec: timing}
\begin{figure}
 \begin{center}
  \FigureFile(80mm,80mm){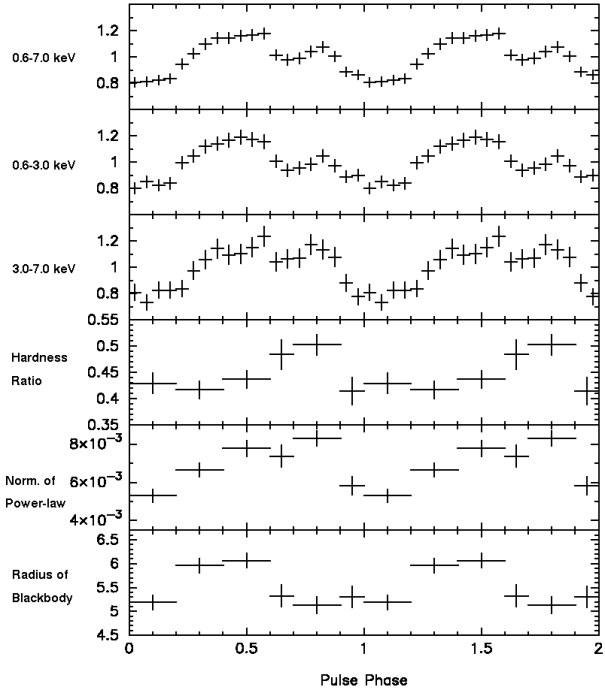}
 \end{center}
  \caption{
	Folded pulse profiles in the energy ranges 
	of 0.6$-$7.0 keV, 0.6$-$3.0 keV, and 3.0$-$7.0 keV
	(top three panels).
	The vertical axes are shown in units of the counts rate
	normalized to the average counts rate.
	The fourth panel shows the hardness ratio,
	the ratio of the photon counts in the hard energy band
	(3.0$-$7.0 keV) to the soft energy band (0.6$-$3.0 keV).
	The bottom two panels show
	the normalization of the power-law and the radius (km) of
	the blackbody emission region on the neutron star surface,
	assuming the absorbed power-law plus blackbody model
	with fixed $N_H$ to the best-fit (Table \ref{table: fit}),
	and the source distance is 7 kpc.
	The horizontal axes are shown in units of pulse phase up to 2 periods.
	The vertical errors in all panels are $1 \sigma$ level.
	In all cases, the backgrounds are subtracted in the way
	described in {\S} \ref{sec: obs and red}.
	}
  \label{fig: pf}
\end{figure}

The event arrival times have been corrected to the value at the
solar system barycenter using {\it axbary} (CIAO v2.2.1),
which utilizes the JPL planetary ephemeris DE-405.
We have investigated the light curves of the pulsar
with binnings of 0.1, 1.0, 4.0, 16.0, and 64.0 s.
We could not find any significant variations of the count rate.
We calculated the pulse period by means of the epoch folding method
using {\it efsearch} v1.1 (XRONOS v5.19).
The resultant pulse period at reference epoch MJD 51,754 is 11.776(1) s
(1$\sigma$ error).
This period is consistent with the spin-down history of the pulsar
(\cite{Gotthelf 2002}).
The background-subtracted pulse profile of 0.6$-$7.0 keV band
is shown in Figure \ref{fig: pf} (the top panel),
where two peaks can been seen clearly
as seen with {\it RXTE} (\cite{Gotthelf 2002}).

The background-subtracted pulse profiles
in the soft energy band (0.6$-$3.0 keV) and hard energy band (3.0$-$7.0 keV)
are also shown in Figure \ref{fig: pf}.
The second peak around phase 0.8 is larger in the harder energy band.
This feature can be seen as an excess of the hardness ratio
(Fig \ref{fig: pf} fourth panel).
This excess is statistically significant
at $99.3\%$ confidence level (C.L.).

The background-subtracted peak-to-peak pulse fraction
($(F_{\rm max} - F_{\rm min})/(F_{\rm max} + F_{\rm min})$)
in the 0.6$-$7.0 keV band, as defined in \citet{Ozel 2001},
is $18.9 \pm 2.6 \%$ ($1 \sigma$ error).

\section{Spectral Analysis} \label{sec: spectral analysis}
The source and background spectra and response files have been generated using
the CIAO v2.2.1 tools {\it dmextract}, {\it mkrmf}, and {\it mkarf}.
The extraction has been performed in pulse-invariant (PI) space.
The spectrum have been grouped into bins, each of those contains
at least 50 events.
All the spectral fits have been limited to the 0.6$-$7.0 keV band and
XSPEC v11.2.0 have been used.

\subsection{Phase-integrated Spectroscopy}
\label{subsec: phase-integrated}
We have fitted the spectrum with a power-law function,
blackbody, and thermal-bremsstrahlung models with
photoelectric absorption $N_H$.
The blackbody and thermal-bremsstrahlung models are
statistically unacceptable at more than 99.9\% C.L.
The power-law function is also unacceptable at 99.2\% C.L.
We have also fitted the spectrum with
power-law plus blackbody model, and
of two blackbodies. Both are acceptable with
$\chi^2/dof$ = 224.6/202 and 225.8/202, respectively
(see Table \ref{table: fit}).

\begin{figure}
 \begin{center}
  \FigureFile(85mm,85mm)
{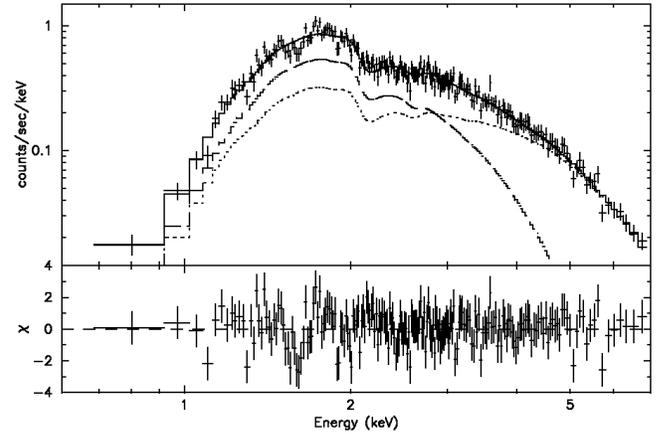}
 \end{center}
  \caption{The upper panel: the background-subtracted pulsar spectrum
	(cross) and the absorbed power-law plus blackbody model spectrum
	  with the best fitting parameters in Table \ref{table: fit}
	 (solid histogram).
	The power-law and blackbody components of this model are shown in
	the dotted and dashed histograms, respectively.
	The bottom panel:  the residual of the
	source counts from the best-fit model.}
  \label{fig: 4}
\end{figure}

The spectrum and the best-fit 
power-law plus blackbody model
are shown in Figure \ref{fig: 4}.
A dip structure around 1.6 keV is a spurious feature
due to the K-edge of aluminum
in the optical blocking filter covering the CCD
({\it Chandra} Help Desk; private communication).

When fitted with power-law plus blackbody model,
the unabsorbed flux in the energy range of 0.6$-$7.0 keV is
$5.0 \times 10^{-11}$ ergs ${\rm s}^{-1}$ ${\rm cm}^{-2}$,
and the blackbody component contributes to 48\% of
the total unabsorbed flux.
Assuming a source distance of $d = 7.0 d_7$ kpc,
X-ray luminosity in the energy range of 0.6$-$7.0 keV is
 $2.9 \times 10^{35} d_7^2$ ergs ${\rm s}^{-1}$, and
the radius of blackbody emission region is
${\rm R}_{\rm BB}= 5.5^{+0.6}_{-0.4} d_7$ km.

When fitted with two blackbodies,
the unabsorbed flux in the energy range of 0.6$-$7.0 keV is
$4.1 \times 10^{-11}$ ergs ${\rm s}^{-1}$ ${\rm cm}^{-2}$,
and the low temperature blackbody component amounts to 75\%
of the total unabsorbed flux.
Assuming a source distance of $d = 7.0 d_7$ kpc,
X-ray luminosity in the energy range of 0.6$-$7.0 keV is
$2.4 \times 10^{35} d_7^2$ ergs ${\rm s}^{-1}$, and
the radii of the blackbody emission regions with the low
and high temperatures are
${\rm R}_{\rm low}= 5.7^{+0.6}_{-0.5} d_7$ km and
${\rm R}_{\rm high}= 0.36^{+0.08}_{-0.07} d_7$ km, respectively.

To compare the flux with the past observations,
we calculate the absorbed and unabsorbed flux
in the energy range of 1.0$-$10.0 keV,
assuming the best-fit absorbed power-law function spectrum.
The results are
$1.9 \times 10^{-11}$ ergs ${\rm s}^{-1}$ ${\rm cm}^{-2}$ and
$6.8 \times 10^{-11}$ ergs ${\rm s}^{-1}$ ${\rm cm}^{-2}$, respectively.
Those values are consistent with {\it ASCA} observations
(\cite{Vasisht Gotthelf 1997}, \cite{Gotthelf Vasisht 1997},
and \cite{Mereghetti 2002}).

\begin{longtable}{cccccc}
  \caption{Phase-Integrated spectrum fitting parameters
	with $1\sigma$ errors. 
	Notation of the models are as follows; PL: Power-law, BB: Blackbody,
	TB: thermal-bremsstrahlung. }
	\label{table: fit}
  \hline\hline
\endfirsthead
  \hline\hline
\endhead
  \hline
\endfoot
  \hline
\endlastfoot
         &  & & & $N_H$ & $\chi^2/dof$ \\
  Model  & $\Gamma$ & $kT$(keV)& $R_{BB}$ (km) & ($10^{22} {\rm cm}^{-2}$) & \\ \hline
  PL  & $3.28 \pm 0.05$ & --- & --- & $3.25 \pm 0.06$ & 255.7/204 = 1.25 \\
  BB  & --- & 0.69      &  2.4   & 1.6                 & 617.7/204 = 3.03 \\
  TB  & --- & $2.01 \pm 0.06 $ & --- & $2.52 \pm 0.04$  & 330.0/204 = 1.62 \\
  PL + BB & $2.0 \pm 0.3$ & $0.44 \pm 0.02$ & $5.5^{+0.6}_{-0.4}$ &
	  $2.54^{+0.15}_{-0.13}$ & 224.6/202 = 1.11 \\ 
  BB + BB & --- & $0.47 \pm 0.02$/$1.5^{+0.2}_{-0.1}$ &
   $5.7^{+0.6}_{-0.5}$/$0.36^{+0.08}_{-0.07}$
          & $2.34 \pm 0.08$ & 225.8/202 = 1.12 \\ \hline

\end{longtable}

\subsection{Phase-resolved Spectroscopy} \label{sec: phase-resolved}
To search for possible spectral variations as a function of
the pulse phase, we have divided the data into six phase intervals:
off-pulse (0.0$-$0.2), rising (0.2$-$0.4),
top of the first pulse (0.4$-$0.6), valley (0.6$-$0.7),
top of the second pulse (0.7$-$0.9), and off-pulse (0.9$-$1.0)
(see Figure \ref{fig: pf}).
We have fitted all phases with absorbed power-law function plus
blackbody model, where the column density $N_H$ at each phase
is fixed to the best value obtained in {\S\S} \ref{subsec: phase-integrated}.
Due to the limited photon statistics,
the fitting with no constraint is not possible.
Therefore, we first fixed only the power-law index,
or blackbody temperature.
In either case, fits for all phases are acceptable within $90 \%$ C.L., and
the blackbody temperatures or the power-law indices
show no significant deviations from the mean values.
Then, we fixed the blackbody temperature
and photon index to the best-fit values
obtained in {\S\S} \ref{subsec: phase-integrated} and normalizations are
allowed to vary. In this case all fits are also acceptable within $90 \%$ C.L.
The variation of the normalizations are shown in Figure \ref{fig: pf}.
The normalization of the blackbody varies in sinusoidal form,
while that of the power-law component does not.
In Figure \ref{fig: pf}, the blackbody normalizations are shown 
as the radii of the emission region on the neutron star surface,
assuming a distance of 7.0 kpc.

We have repeated the similar procedure for the
absorbed two blackbody model.
We obtained essentially the same result with the previous model,
where the normalization of the power-law component is replaced with
the higher temperature blackbody component
in the bottom panels of Figure \ref{fig: pf}.

\section{Spatial search for extended nebula} \label{sec: nebula}
We fit the one-dimensional CC mode image
with a model consisting of a point source (pulsar)
and a constant background (SNR), and evaluated the normalizations
of the two components.
To simulate the point source,
we used {\it MARX} v4.0 and input
the best-fit power-law plus blackbody model spectrum,
obtained in {\S\S} \ref{subsec: phase-integrated}.
The model with the best-fit normalizations is shown
in Figure \ref{fig: 1d}.
The residual from the model is largest at $\pm 0.5$ arcsec, indicating
that the image is more extended than the point spread function
by $\sim 0.5$ arcsec ($\approx 1$ pixel).
We, however, do not consider this spatial extent to be significant,
because the image is constructed with dithering and rotation,
and is unavoidably blurred to an order of 1 pixel.

We then estimated the upper limits for a possible pulsar nebula flux.
We fit the image with a model including a point source,
extended nebula, and constant background.
The nebula is simulated by {\it MARX} v4.0
as a sphere filled with uniform emissivity centered
on the point source, and the spectrum is the absorbed power-law
model with a photon index of $\Gamma = 2.0$
and $N_H = 2.54 \times 10^{22} {\rm cm}^{-2}$.
We obtained upper limits for the nebula with radii of 1.0 and 3.0 arcsec
to be 
$3.0 \times 10^{-12}$ ergs ${\rm s}^{-1}$ ${\rm cm}^{-2}$ and
$3.7 \times 10^{-13}$ ergs ${\rm s}^{-1}$ ${\rm cm}^{-2}$
in the energy range of 0.6$-$7.0 keV,
respectively.

\section{Discussion}
Thanks to the unprecedented spatial resolution of {\it Chandra},
the pulsar spectrum is clearly discriminated from the surrounding SNR
for the first time.
Before this observation,
this source is only fitted with a power-law function,
then resulting in the power-law index of
$\Gamma = 3.4 \pm 0.3$ 
by {\it ASCA} (0.9$-$8.0 keV) \citep{Gotthelf Vasisht 1997}
and $\Gamma \simeq 2.0$ by {\it RXTE} (2$-$10 keV)
\citep{Gotthelf 2002}.

In our analysis, we find that the power-law
plus blackbody model is preferred to a single
power-law function, where
the best spectral parameters are
$\Gamma = 2.0 \pm 0.3$, $kT_{\rm BB} = 0.44 \pm 0.02$ keV,
and $N_H = 2.54^{+0.15}_{-0.13} \times 10^{22} {\rm cm}^{-2}$.
This spectral model may be typical for AXPs,
because four AXPs, 1E 2259+586, 4U 0142+61, 1E 1048.1-5937, and
1RXS J170849.0-400910 are also fitted with this model
(e.g. \cite{Mereghetti 2002}).
The blackbody temperature of 1E 1841-045 is similar to these four AXPs,
but the power-law index is significantly flatter
than that of others ($\gtrsim 3$).
This value is close to that of SGRs ($\sim 2$),
and its estimated dipole magnetic field
($B = 3.3 \times 10^{19}({\rm P} \dot{\rm P})^{1/2} =
 7.3 \times 10^{14}$ G) is also close to that of SGRs
(\cite{Marsden White 2001});
$5.9 \times 10^{14}$ G for SGR 1900+14 and 
$9.7 \times 10^{14}$ G for SGR 1806-20.
From these aspects, 1E 1841-045 is the closest object
to SGRs among AXPs.

The radius of the blackbody emission region is
$5.5^{+0.6}_{-0.4} d_7$ km.
This is smaller than the whole neutron star surface,
but yet much greater than
the polar cap hot spot expected for
magnetically channeled accretion.
Furthermore, the blackbody radius varies with pulse phases
in sinusoidal form, suggesting the inhomogeneous 
distribution of the temperature on the neutron star surface,
which might be caused by non-isotropic heat conduction
(\cite{Greenstein Hartke 1983}), or
heating due to magnetic field decay.

Spectral hardening at the second peak is found for the first time.
This feature could be attributed to
an excess of the power-law component at that phase,
when fitted with a power-law plus blackbody model.
Since the phase of maximal normalization of the power-law component is
different from the blackbody maximum,
the emission region of the power-law component is
clearly displaced from the hot thermal region.
We also obtained a good fit for a two-blackbody model.
Therefore, we can interpret those that
there are two emission regions with different energy spectra;
one is a hot spot on the neutron star surface
which emits lower temperature blackbody radiation,
and the other is a non-thermal source 
presumably in the magnetosphere, or
another hot spot on the neutron star surface with a higher temperature.

\end{document}